\documentclass[sigconf,nonacm]{acmart}

\usepackage{url}

\usepackage{booktabs} 

\usepackage{subfigure}
\usepackage{amsmath}
\usepackage{graphicx}
\usepackage{enumitem} 
\usepackage{color}

\begin{document}

\title{A Systematic Media Frame Analysis of 1.5 Million New York Times Articles from 2000 to 2017
\subtitle{[Please cite the WebSci'20 version of this paper]}
}

\author{Haewoon Kwak}
\affiliation{%
  \department{Qatar Computing Research Institute}
  \institution{Hamad Bin Khalifa University}
  \city{Doha}
  \country{Qatar} 
}
\email{haewoon@acm.org}

\author{Jisun An}
\affiliation{%
  \department{Qatar Computing Research Institute}
  \institution{Hamad Bin Khalifa University}
  \city{Doha}
  \country{Qatar} 
}
\email{jisun.an@acm.org}

\author{Yong-Yeol Ahn}
\affiliation{%
  \department{Indiana University, Bloomington}
  \city{IN}
  \country{USA} 
}
\email{yyahn@iu.edu}

\begin{abstract}
Framing is an indispensable narrative device for news media because even the same facts may lead to conflicting understandings if deliberate framing is employed.  
Therefore, identifying media framing is a crucial step to understanding how news media influence the public.
Framing is, however, difficult to operationalize and detect, and thus traditional media framing studies had to rely on manual annotation, which is challenging to scale up to massive news datasets. 
Here, by developing a media frame classifier that achieves state-of-the-art performance, we systematically analyze the media frames of 1.5 million New York Times articles published from 2000 to 2017. 
By examining the ebb and flow of media frames over almost two decades, we show that short-term frame abundance fluctuation closely corresponds to major events, while there also exist several long-term trends, such as the gradually increasing prevalence of the ``Cultural identity'' frame. 
By examining specific topics and sentiments, we identify characteristics and dynamics of each frame. 
Finally, as a case study, we delve into the framing of mass shootings, revealing three major framing patterns. 
Our scalable, computational approach to massive news datasets opens up new pathways for systematic media framing studies. 
\end{abstract}




\maketitle

\section{Introduction}

News media influence our understanding of, our beliefs about, and  our attitudes toward what is happening around us~\cite{entman1989media}. 
On the one hand, by selecting \emph{what to report}\textemdash which is not always aligned with what people want to know~\cite{kwak2018we}\textemdash news media can create biased public awareness (\textit{``agenda-setting''})~\cite{mccombs1972agenda}. 
For instance, the proportion of Gallup poll respondents who called the Gulf crisis the most important national problem could be simply explained by the amount of its media coverage~\cite{iyengar1993news}. 
On the other hand, by determining \emph{how to report} (\textit{``framing''}), news media can potentially alter public attitudes on particular issues~\cite{lakoff2014all,chong2007framing,nelson1997media,baumgartner2008decline,tan2018you}.
Framing is implemented by emphasizing certain aspects of an issue more than other aspects and making them more  salient~\cite{chong2007framing,entman1993framing}.
For example, when reporting on the issue of poverty, some news media may focus on unemployed individuals, while others may focus on national policies.
Such differences in focus have been shown to be pivotal; according to one study~\cite{iyengar1994anyone}, those who are exposed to the former framing of poverty were more likely to blame it on individual failings, while those exposed to the latter were more likely to blame it on the government or other forces beyond their control. 
Therefore, to understand their influence on the public, it is crucial to study framing\textemdash in addition to the agenda setting\textemdash in news media.

In contrast to the agenda-setting and selection bias, which have been extensively studied~\cite{iyengar1993news,saez2013social,kwak2014first}, media framing has posed challenges to researchers because of the inherent complexity and vagueness of the concept. 
Media frames are difficult to concretely operationalize as they are often abstract and subtle~\cite{matthes2008content,boydstun2014tracking}.
Most of the studies on media frames are thus conducted based on manual labels obtained for a handful of specific issues and  issue-specific frames, such as ``global warming'' vs. ``climate change''~\cite{schuldt2011global}, ``gay civil unions'' vs. ``homosexual marriage''~\cite{price2005framing}, ``punishment'' vs. ``innocence'' (in relation to the death penalty)~\cite{baumgartner2008decline}, and more~\cite{schnell2001assessing}.
Finding a systematic approach to media framing that transcends these  issues remains a critical challenge.

A prominent trend in the research that aims to address this limitation is the annotation of a large corpus of news articles with a standardized set of media frames that are universal across multiple issues.
Probably the most notable such effort is the ``Media Frames Corpus'' (MFC)~\cite{card2015media}, which labels 20,037 news articles from 13 U.S. national newspapers on the topics of  immigration, smoking, and same-sex marriage with one of 15 topic-agnostic general media frames (shown in Table~\ref{tab:general_frames})~\cite{boydstun2014tracking}.

\begin{table*}[h!]
\footnotesize \frenchspacing
\begin{center}
  \begin{tabular}{p{2.3cm}p{5cm}p{9cm}}
    \toprule
     Frame & Short description & Over-representative words in our dataset computed by Eq. (3) \\
    \toprule
    
    Capacity and resources & Availability of physical, human or financial resources, and capacity of current systems & computer, web, airport, www, water, trains, service, available, passengers, transportation, flights, agency, number, delays, applications, airports, software, transit, site, system \\
    \midrule
    
    Crime and punishment & Effectiveness and implications of laws and enforcement & police, prosecutors, charges, officers, arrested, prison, charged, guilty, officer, criminal, convicted, pleaded, authorities, investigation, sentenced, crime, murder, arrest\\
    \midrule
    
    Cultural identity & Traditions, customs, or values of a social group in relation to a policy issue & theater, org, street, 212, art, through, museum, game, music, season, play, saturdays, sundays, show, gallery, 30, avenue, film, arts, exhibition \\
    \midrule
    
    Economic & Costs, benefits, or other financial implications & percent, company, billion, companies, market, million, investors, prices, tax, stock, sales, financial, business, price, bank, its, investment, revenue, economy, growth \\ \midrule
    
    External regulation and reputation & International reputation or foreign policy of the U.S. & pm, united, iran, nations, nuclear, korea, states, iraq, russia, israel, china, am, countries, minister, military, palestinian, weapons, north, foreign, administration\\
    \midrule
    
    Fairness and equality & Balance or distribution of rights and responsibilities & editor, article, writer, editorial, op, rights, discrimination, ed, readers, aug, civil, racial, our, freedom, equality, gay, is, right, column, equal\\
    \midrule
    
    Health and safety & Health care, sanitation, public safety & dr, patients, disease, researchers, health, study, medical, cancer, doctors, drug, cells, medicine, scientists, patient, drugs, brain, virus, treatment, blood, hospital \\
    \midrule
    
    Legality, constitutionality and jurisprudence & Rights, freedoms, and authority of individuals, corporations, and government & court, judge, justice, lawyers, case, supreme, ruling, appeals, law, legal, lawyer, trial, lawsuit, justices, federal, filed, courts, plaintiffs, judges, decision\\
    \midrule
    
    Morality & Religious or ethical implications & church, catholic, bishops, religious, pope, vatican, bishop, priests, cardinal, religion, christian, archbishop, god, catholics, rev, faith, christians, episcopal, jesus, gay\\
    \midrule
    
    Policy prescription and evaluation & Discussion of specific policies aimed at addressing problems & feedback, essentials, interested, confirm, prior, tell, page, your, cooking, purchase, below, regulations, us, rules, think, proposal, ban, commission, environmental, zoning\\
    \midrule
    
    Political & Considerations related to politics and politicians, including lobbying,  elections, and voters & republican, mr, democrats, republicans, campaign, senate, democratic, senator, party, voters, election, obama, bush, vote, clinton, political, candidates, governor, president, candidate \\
    \midrule 
    
    Public opinion & Attitudes and opinions of the general public, including polling and demographics & protesters, protests, protest, demonstrators, points, rally, poll, saturday, sunday, organizers, derby, scored, demonstrations, opposition, crowd, yards, activists, game, victory, park  \\
    \midrule
    
    Quality of life & Threats and opportunities for the individual's  wealth, happiness, and well-being & her, she, my, mother, father, he, his, me, family, daughter, husband, wife, son, was, school, friends, friend, beloved, graduated, life  \\
    \midrule
    
    Security and defense & Threats to welfare of the individual, community, or nation & shorefront, comers, privatization, homeowners, asks, military, qaeda, attacks, security, al, forces, iraqi, officials, opinion, army, attack, intelligence, soldiers, iraq, land \\
    \bottomrule
  \end{tabular}
    \caption{Overview of general media frames. Short descriptions are from \cite{card2015media}. The third column lists the overrepresentative keywords associated with articles of each frame ranked by log-odds ratio. We omitted `Other' category. }
    \label{tab:general_frames}
    \end{center}
\end{table*}

Here, leveraging the MFC, we develop a general media frame classifier and apply it to the near-complete set of news articles published in the New York Times between January 2000 and December 2017, aiming to understand the temporal and system-wide patterns. 
The New York Times archive has been a subject of many studies~\cite{druckman2001limits,chyi2004media,rauch2007seattle} because it does not only exhibit one of the largest digitized news article databases yet available but also is one of the most prominent agenda-setting media outlets, widely read by the public and policy makers~\cite{gitlin2003whole,muschert2006media,kothari2010framing}.

Our media frame classifier enables us to investigate media frames of the large-scale news corpus that cannot be manually analyzed due to its scale. In contrast to previous studies focusing on news about particular issues over shorter periods, our study infers general media frames from 1.5M news articles over 18 years, allowing us to reveal longitudinal patterns and general characteristics of media framing across different issues. 
Specifically, we address the following research questions:
\begin{itemize}[leftmargin=*]
    \item[] \textbf{RQ1:} What are the long-term trends of media framing? 
    \item[] \textbf{RQ2:} What are the dynamics of framing at the level of individual issues?
    \item[] \textbf{RQ3:} How is each frame delivered? Is there a specific linguistic style (e.g., sentiment)?
    \item[] \textbf{RQ4:} (Case study) What frames have been used to report different mass shootings in the United States?    
\end{itemize}

\section{Data Collection}

Using the ``Fake News Corpus''\footnote{https://github.com/several27/FakeNewsCorpus}, we collected 1.5 million New York Times articles (about 8,500 articles per month) published between January 1, 2000 and December 31, 2017. 
As the Fake News Corpus is built for fake news recognition, to ensure the lack of systematic bias, we first check the coverage of our dataset by using the New York Times Archive API\footnote{\url{https://github.com/NYTimes/public_api_specs/blob/master/archive_api/archive_api.md}; of the four types of documents \textemdash article, audio, blogpost, and multimedia \textemdash we only consider ``articles.''}.
Overall, our dataset covers 99.38\% of New York Times articles published from January 1, 2000 to December 31, 2017; the lowest monthly coverage is 95.38\% in February 2013. 
In other words, it is a near-complete collection of the New York Times articles published between 2000 and 2017.

\section{Media Frame Classifier}

To estimate the prevalence of each media frame, we first build an article-level media frame classifier. 
It is trained to predict the primary frame of an individual article using the labeled Media Frames Corpus (MFC)\footnote{\url{https://github.com/dallascard/media_frames_corpus}}.
We used the state-of-the-art language representation model, BERT (Bidirectional Encoder Representations from Transformers)~\cite{devlin2019bert}. 
We fine-tuned the pre-trained BERT-base model\footnote{\url{https://github.com/google-research/bert}}  
with our training data using a small learning rate, 0.0002, a maximum sequence number of 128, a batch size of 32, and a number of training epochs of 3. 
Since the MFC dataset is imbalanced, we added class weights in the loss function. 
The performance of the model, trained on 11k articles, is F$_1$=71.34, based on a test set of 1,138 articles. 
The best accuracy yet reported among similar approaches using the same corpus, was 58.4\% on the ``Immigration'' subset, in the previous study by Ji and Smith (\citeyear{ji2017tree}), which is excelled by our presented method.

In addition to the quantitative evaluation, we present  qualitative validation of the identified frames using the keywords of the articles. 
New York Times articles include \texttt{META} keywords in the form of \texttt{HTML} head tags for search engine optimization (SEO).
For example, a news article titled ``Yeltsin Resigns: In Moscow; Yeltsin Leaves Russians Glad or Uninterested''\footnote{https://www.nytimes.com/2000/01/01/world/yeltsin-resigns-in-moscow-yeltsin-leaves-russians-glad-or-uninterested.html} has 
\texttt{$<$meta name=``keywords'' content=``Russia,Yeltsin Boris N, Putin Vladimir V, Sus\-pensions dismissals and resignations, Politics and gove\-rnment''/$>$} as its HTML head tags.
Table~\ref{tab:1_2_frame} shows the top 10 most frequent META keywords and their top two most frequent frames. 
We also present the prevalence of the most frequent frame.

We can see intuitive agreements between the top META keywords and the detected frames.
For example, the Cultural identity and Quality of life frames are frequently found in articles with culture-, art-, and entertainment-related keywords, such as Books and literature, Music, Baseball, and Reviews. The Security and defense frame is found in articles about Terrorism, and the Political frame in  articles with the keywords Politics and government and United States politics and government.

\begin{table}[t!]
\frenchspacing
\begin{center}
  \begin{tabular}{p{0.5cm}p{2cm}p{4.2cm}}
    \toprule
    Rank & Keyword & Frequent media frames \\
    \toprule
    1 & New York City & Cultural identity (20.8\%), Quality of life \\
    \midrule
    2 & Books and literature & Cultural identity (35.0\%), Quality of life\\
    \midrule
    3 & Terrorism & Security and defense (30.2\%), External regulation\\
    \midrule
    4 & US international relations & External regulation (46.2\%), Security and defense \\
    \midrule
    5 & Music & Cultural identity (56.4\%), Quality of life \\
    \midrule
    6 & Baseball & Cultural-identity (58.3\%), Quality-of-life \\
    \midrule
    7 & Computers \& Internet & Economic (38.1\%), Cultural-identity \\
    \midrule
    8 & Reviews & Cultural-identity (57.8\%), Quality-of-life \\
    \midrule
    9 & Politics and government & Political (45.7\%), External-regulation \\
    \midrule
    10 & US politics and government & Political (55.3\%), Economic \\
    \bottomrule
  \end{tabular}
    \caption{The two most frequent frames in news articles for the top 10 META keywords}
    \label{tab:1_2_frame}
    \end{center}
\end{table}

\begin{table}[h!]
\frenchspacing
\begin{center}
  \begin{tabular}{p{1.3cm}p{6.4cm}}
    \toprule
    Section & Frequent media frames \\
    \toprule
    Sports & Cultural identity (50.1\%), Quality of life \\
    \midrule
    Business & Economic (68.1\%), Cultural identity\\
    \midrule
    World & External regulation (25.7\%), Security and defense\\
    \midrule
    Opinion & Political (16.0\%), Economic \\
    \midrule
    Arts & Cultural identity (58.8\%), Quality of life \\
    \bottomrule
  \end{tabular}
    \caption{The two most frequent frames in the top 7 news sections}
    \label{tab:section_1_2_frame}
    \end{center}
\end{table}

Similarly to the keyword-level analysis, we now compare frequent frames between news sections. New York Times articles embed sections in a full URL. For example, the URL  \url{https://www.nytimes.com/2020/02/21/health/coronavirus-cases-usa.html} shows that the article appears in the `Health' section. 
Table~\ref{tab:section_1_2_frame} shows the two most frequent frames in the top five news sections, including the prevalence of the most frequent frame.  
We note that the `Sports' section (1st) has 146,732 articles, and the `Arts' section (5th) has 91,364 articles. 
While there are no ground-truth labels to verify the result directly, the two most frequent frames in each section closely  reflect the general characteristics of that section. 
Also, while the Politics section is not among the top 5 sections, the Political frame is the most prevalent (63.1\%) in that section. 
The results of two qualitative evaluations to identify frames are thus closely aligned with keywords and sections, indicating that our classifier performs reasonably well and can be adopted for  further analyses.

\subsection{Statistically Over-represented Frame Words}
To characterize articles of each frame, we investigate unigrams that are over-represented in each frame. To reliably compute the over-representation, we use log-odds ratios with informative Dirichlet priors~\cite{monroe2008fightin}.
This method estimates the log-odds ratio of each word $w$ between two corpora $i$ and $j$ given the prior frequencies obtained from a background corpus $\alpha$. The log-odds ratio for word $w$, $\delta_w^{i-j}$ is estimated as:

\begin{equation}\label{eq:delta}
    \delta_w^{(i-j)} = \log\frac{y_w^i + \alpha_w}{n^i + \alpha_0 - y_w^i - \alpha_w} - \log\frac{y_w^j + \alpha_w}{n^j + \alpha_0 - y_w^j - \alpha_w}
\end{equation}
\noindent where $n^i$ (resp. $n^j$ ) is the size of corpus $i$ (resp. $j$), $y_w^i$  (resp. $y_w^j$) is the count of word $w$ in corpus $i$ (resp. $j$), $\alpha_0$  is size of the background corpus, and $\alpha_w$  is the frequency of word $w$ in the background corpus.
Furthermore, the method provides an estimate for the variance of the log-odds ratio and $z$-score, as follows: 
\begin{equation}\label{eq:sigma}
    \sigma^2(\delta_w^{(i-j)}) \approx \frac{1}{y_w^i + \alpha_w} + \frac{1}{y_w^j + \alpha_w},
\end{equation}
\begin{equation}\label{eq:zscore}
    Z = \frac{\delta_w^{(i-j)}}{\sqrt{\sigma^2(\delta_w^{(i-j)})}}
\end{equation}

We identify words associated with each frame with an exploratory analysis using the log-odds ratio. To remove noisy estimates and rare words, we select words that appear at least 100 times in the entire corpus.
We then extract all unigrams from the news articles and compute their log-odds ratio using Eq.~(\ref{eq:delta}). For the prior, background word frequency is computed from the entire New York Times dataset, rather than the sample. 
The words are then ranked by their estimated $z$-scores, computed using Eq.~(\ref{eq:zscore}).
The 20 most over-represented unigrams for each media frame are presented in Table~\ref{tab:general_frames}.

\begin{figure*}[ht]
\centering
\includegraphics[width=\textwidth]{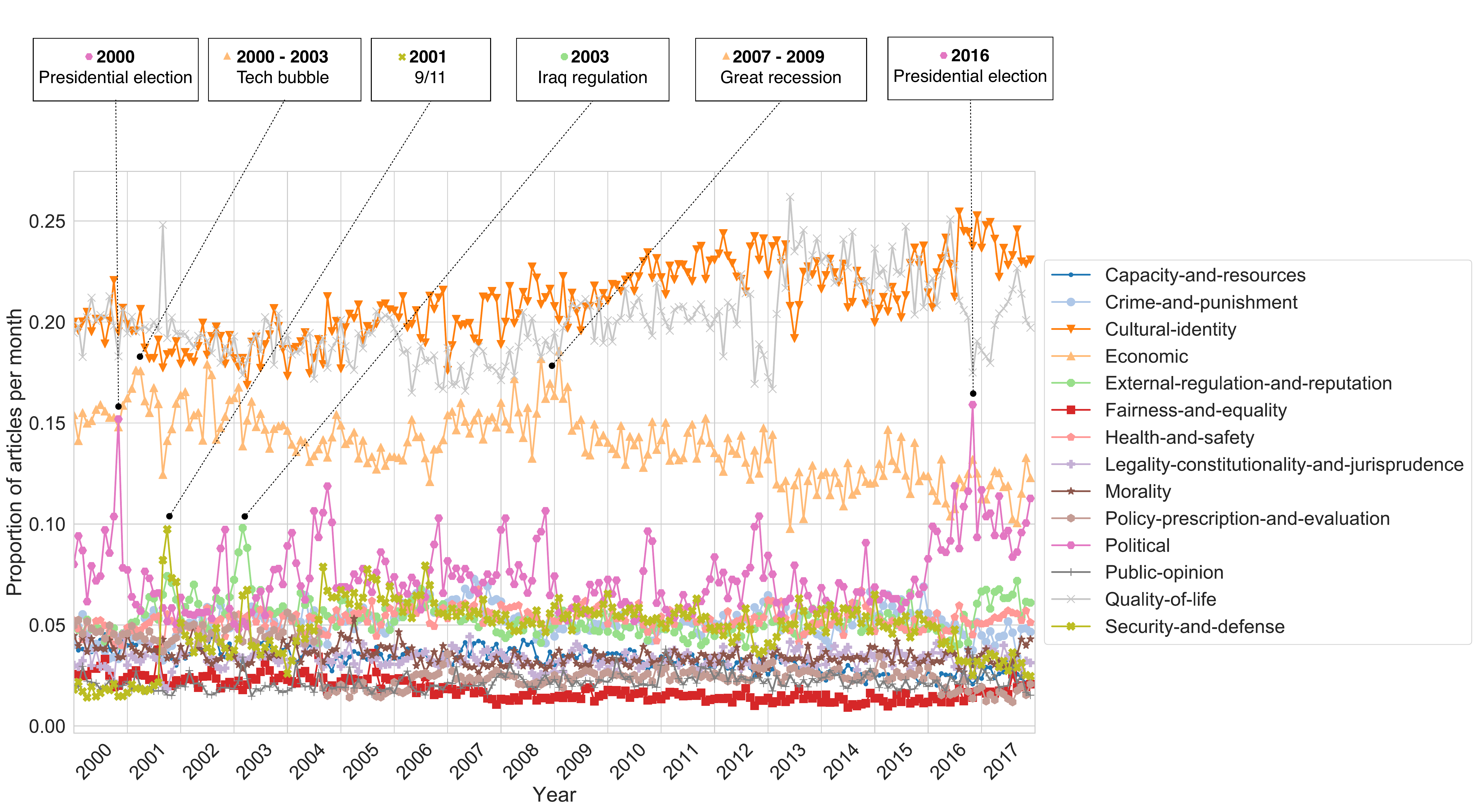}
\caption{[Zoomable in PDF] Prevalences of articles per frame over time
}
\label{fig:narticles_overtime_perframe}
\end{figure*}

The top five words for the Crime and punishment frame are \textit{police, prosecutors, charges, officers,} and \textit{arrested}. 
For the Cultural identity frame, culture-related words like \textit{theater, museum, game,} and \textit{music} are shown in the table, while for Quality of life, family-related words like \textit{mother, fathers, family, beloved,} and \textit{life} are appeared. 
For the Fairness and equality frame, the top five words are \textit{editor, article, writer, editorial,} and \textit{op}, indicating that this frame tends to appear in Opinion articles in the New York Times. 
Overall, the over-represented words show a strong association with their corresponding frame. This improves our understanding of how  different media frames are used in news reporting.

\section{Long-term Framing Trends}

In this section, we answer our first research question, \textit{RQ1: What are the long-term trends of framing?} 
We examine the temporal abundance of each frame across all news articles published in the New York Times in the period under investigation. As each article is assigned a single media frame, we simply calculate the monthly average fractions of each frame and tack them (the ``Other'' frame is excluded).

Figure~\ref{fig:narticles_overtime_perframe} shows the evolution of each frame over time.
The first long-term trend that we can observe is the consistent prominence of the Cultural identity (orange), Quality of life (gray), and Economic (light orange) frames, as well as the gradual shift from economic- to culture- and quality of life-related frames. 
This long-term trend may be a reflection of the general societal trend toward more sophisticated cultural needs~\cite{grossberg2006mediamaking} and market forces~\cite{hamilton2004all} over time. 
The Economic frame, while slowly declining, also reflects major economic events in the U.S. 
The higher prevalence of the Economic frame in the periods from  early 2000 to 2003 matches the bursting of the tech bubble in 2000 (and subsequent events such as the 9/11 attacks in 2001 and the military conflicts of 2002-3); the Economic frame became prevalent again with the bursting of the housing bubble and the Great Recession (from 2007 to 2009). 
The decreased prevalence afterwards may reflect the gradual recovery of the economy (and the corresponding loss of focus on this field) since the crisis. 

The Political frame (purple) fluctuates with the election cycle. Its two strongest surges are in 2000 and 2016, coinciding with the two presidential elections that elected Republican presidents and created a lot of controversy. 
It is also interesting to see that the proportion for the Political frame stays high after the 2016 presidential election, indeed higher than before; the monthly average of in the 12 months from January 2017 (0.100) is higher than for the 12 months to September 2016 (0.089), and the difference is statistically significant by the  Mann-Whitney U Test (U=39.0, $p=0.030$).
The other prominent peak corresponds to the Iraq War. 
We observe a surge in the Security and defense frame in 2001 after the September 11 attacks, but overall it has become less prominent over time. 

The External regulation frame surged during two time periods: from 2001 till mid-2003, probably due to the September 11 attacks, and after 2017, due to the Muslim travel ban and the strengthening of immigration regulation in the U.S. 
The Policy prescription frame formerly comprised 5\% of all articles on average, but this suddenly dropped to 2.5\% in late 2004  and has shown little change since then. 
The remaining frames show a low but consistent level over time; on average, the Morality, Legality, Health, Crime, and Public opinion frames are included in 4.2\%, 4.4\%, 5.1\%, 5.2\%, and 2.6\% of all articles, respectively. 
 
In summary, the four most prominent frames \textemdash Cultural identity, Quality of life, Economic, and Political \textemdash compete with each other, and their historical prevalence reflected major social events. Furthermore, the Cultural identity frame steadily increases in prevalence.

While the total prevalence of each frame is fairly stable over time, this does not necessarily mean that the \emph{issues} or the \emph{topics} are also stable. 
To examine the temporal evolution of frames in relation to different topics, we extract the most representative words for each frame over time by using the log-odds ratio.  
We first extract articles representing each frame published in each year. We then extract all unigrams from each one-year sample and compute their log-odds ratios against the all-other-years sample using Eq.~(\ref{eq:delta}). 
For the prior, background term frequency is computed for all articles relating to the corresponding frame. 
The terms are then ranked by their estimated $z$-scores, computed using Eq. (\ref{eq:zscore}).

\begin{table}[th!]
\footnotesize \frenchspacing
\begin{center}
  \begin{tabular}{p{0.3cm}p{3.6cm}p{3.6cm}}
    \toprule
    \textbf{Year} & \textbf{Economic} & \textbf{Security} \\
    \midrule
    2001 & attacks, lucent, yesterday, sept, giuliani, cipro, anthrax, terrorist, vallone, swissair, bush, cavallo, difrancesco, compaq, dynegy & laden, bin, anthrax, alliance, hijackers, sept, taliban, trade, today, osama, b6, mazar, hanssen, center, hijacked \\
    \midrule
    2005 & bathrooms, purcell, mci, katrina, refco, unocal, hurricane, bayou, cnooc, est, guidant, starr, bedrooms, orleans & should, orleans, beach, homeowners, privatization, comers, shorefront, asks, opinion, debate, hurricane, room, allowed, open, mehlis \\
    \midrule
    2010 & bp, spill, photo, goldman, greece, banks, haiti, ipad, recovery, gulf, deepwater, obama, paladino, crisis, renminbi & bp, wikileaks, marja, afghan, should, shahzad, rig, spill, galea, blowout, shorefront, comers, privatization, homeowners, asks,\\
    \midrule
    2015 & eurozone, greece, tsipras, valeant, greek, puerto, rico, rubio, website, kaisa, varoufakis, dorsey, volkswagen, syriza, mylan & islamic, state, paris, isis, houthis, que, isil, syria, migrants, abaaoud, kouachi, hebdo, hungary, kulluk, hollande\\
    \midrule
    2017 & trump, uber, kalanick, bitcoin, kushner, hna, obamacare, snap, tax, mnuchin, wsj, equifax, columnists, amazon, tech & trump, duterte, marawi, uber, rohingya, mattis, inbox, photo, irma, your, twitter, hacking, briefing, macron, korea \\
    \bottomrule
  \end{tabular}
    \caption{Top 15 keywords associated with articles incorporating the Economic and Security and defense frames over time. For example, ``isis'' and ``jihadists'' emerge strongly in 2015.}
    \label{tab:keywords_byframe_overtime}
    \end{center}
\end{table}

We present the top 15 most representative words in the Economic and Security \& defense frames for a selected time period in Table~\ref{tab:keywords_byframe_overtime}.
The words in each year clearly show the issues presented with the corresponding frame. 
For example, `giuliani', mayor of New York City during 9/11, and `dynegy', a company that filed for bankruptcy protection, appear in 2001, `katrina', `orleans', and `hurricane' in 2005, `goldman', `greece', `banks', and `ipad' in 2010, `eurozone', `puerto rico', and `volkswagen' in 2015, and `bitcoin', `obamacare', `equifax', and `amazon' in 2017. 
The hurricane-related words in 2005 support previous findings that cost is one of the most important dimensions in reporting on natural disasters~\cite{houston2012disaster}.

The Security frame also reveals relevant keywords, for example,  `taliban', `hijacked', and `osama bin laden' in 2001, `orleans' and `hurricane' in 2005, `wikileaks', `bp', and `spill' in 2010, `islamic', `paris',  `isis', and `hebdo' in 2015, and `marawi', `rohingya', `hacking', and `korea' in 2017. 
Conflicts in the Middle East, North Korea, and the Philippines, and recent privacy issues around Facebook are framed as threats to the United States (e.g., the U.S. supports the Philippine Government around Marawi conflicts) and thus relate to the Security frame.

Some issues can be associated with multiple frames because they are inherently multidimensional~\cite{schnell2001assessing}. 
For example, the 9/11 attacks in 2001 and Hurricane Katrina in 2005 had a massive impact, and thus can be expected to be reported with various frames. 
Our results show an interesting contrast between  the two frames for a news event. 
After the 9/11 attacks, the Economic articles tend to provide more reporting on companies (e.g., `swissair' or `dynegy'). In contrast, the Security and defense articles focus more on the hijackers and their affiliates.
Likewise, we see that Hurricane Katrina not only had massive economic impact (e.g., the value of houses with three `bedrooms'), but also naturally brought up the security of the nation against  natural disasters (e.g., `homeowners' and `privatization').

\begin{figure*}[h!]
\centering
\subfigure[Legend]{\includegraphics[width=0.14\textwidth]{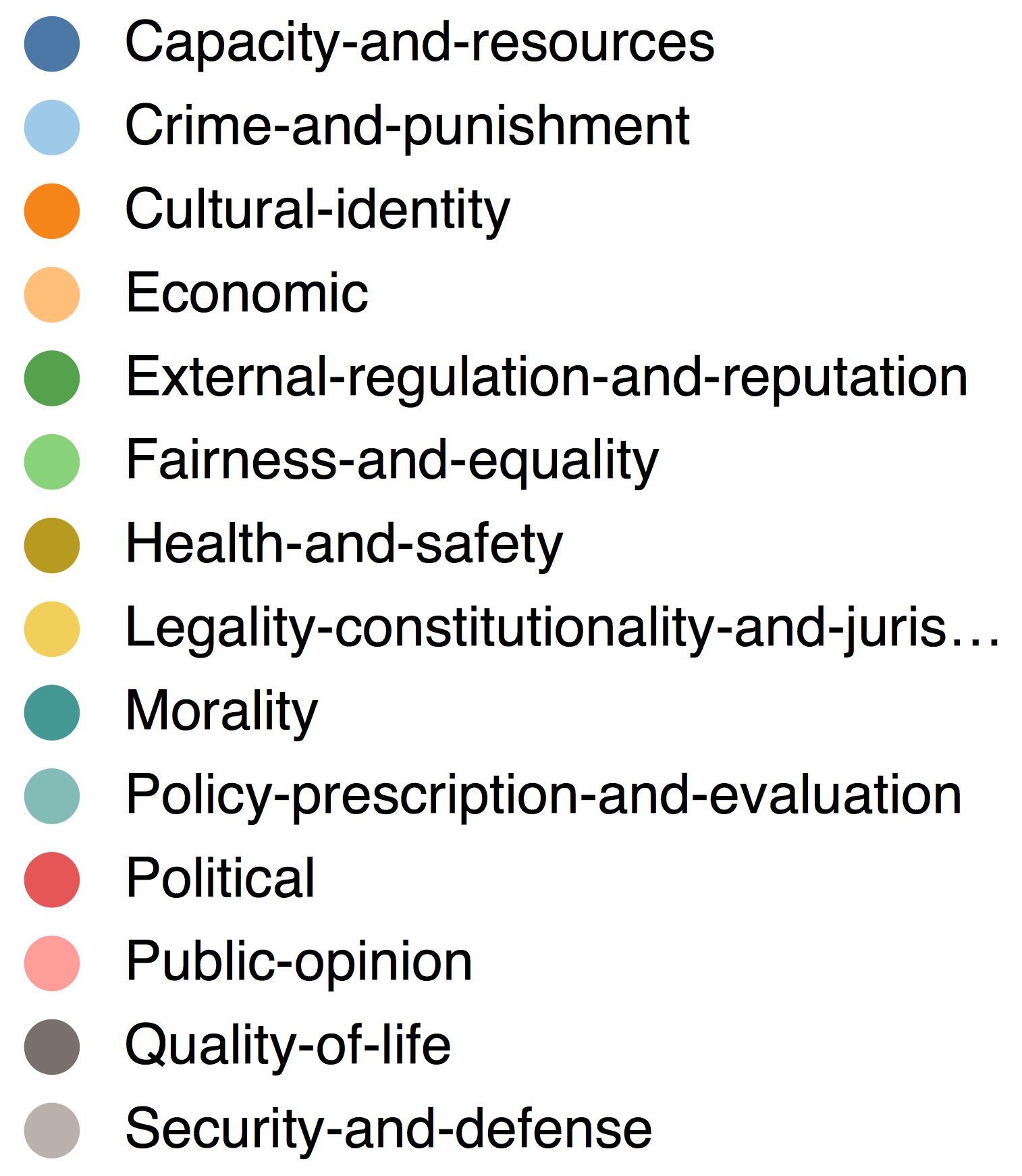}}
\subfigure[Abortion]{\includegraphics[width=0.23\textwidth]{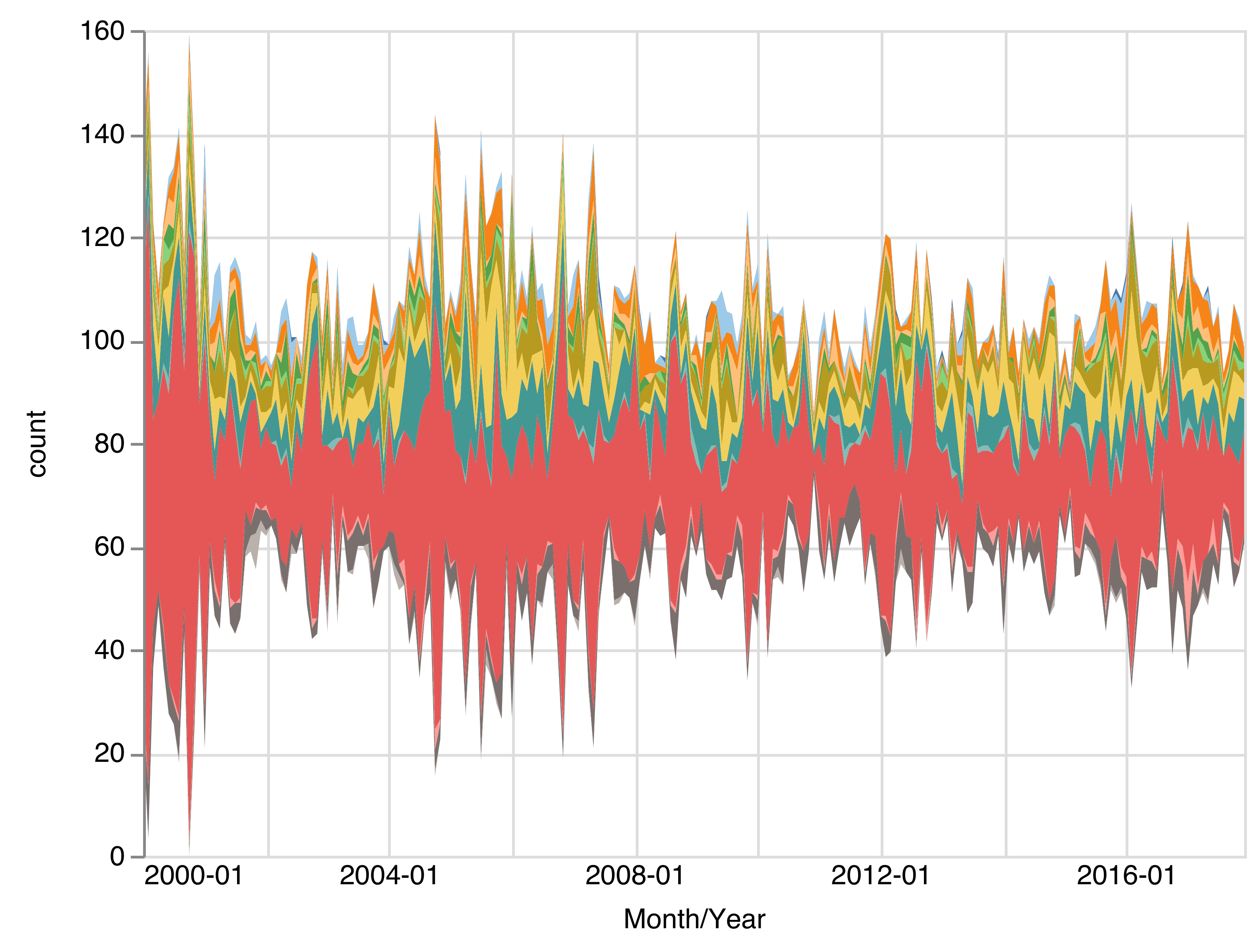}\label{fig:abortion}}
\subfigure[Smoking]{\includegraphics[width=0.23\textwidth]{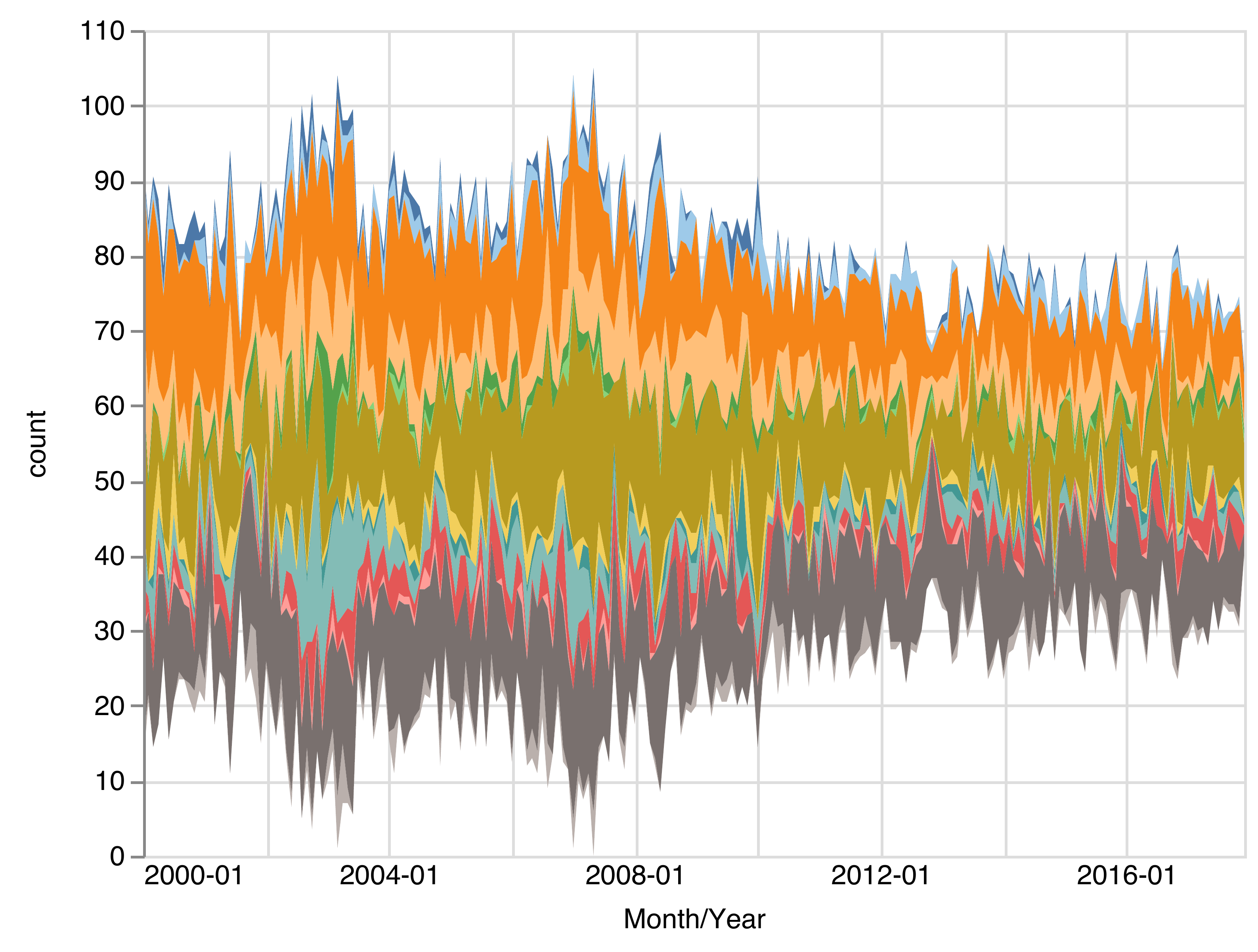}\label{fig:smoking}}
\subfigure[Immigration]{\includegraphics[width=0.23\textwidth]{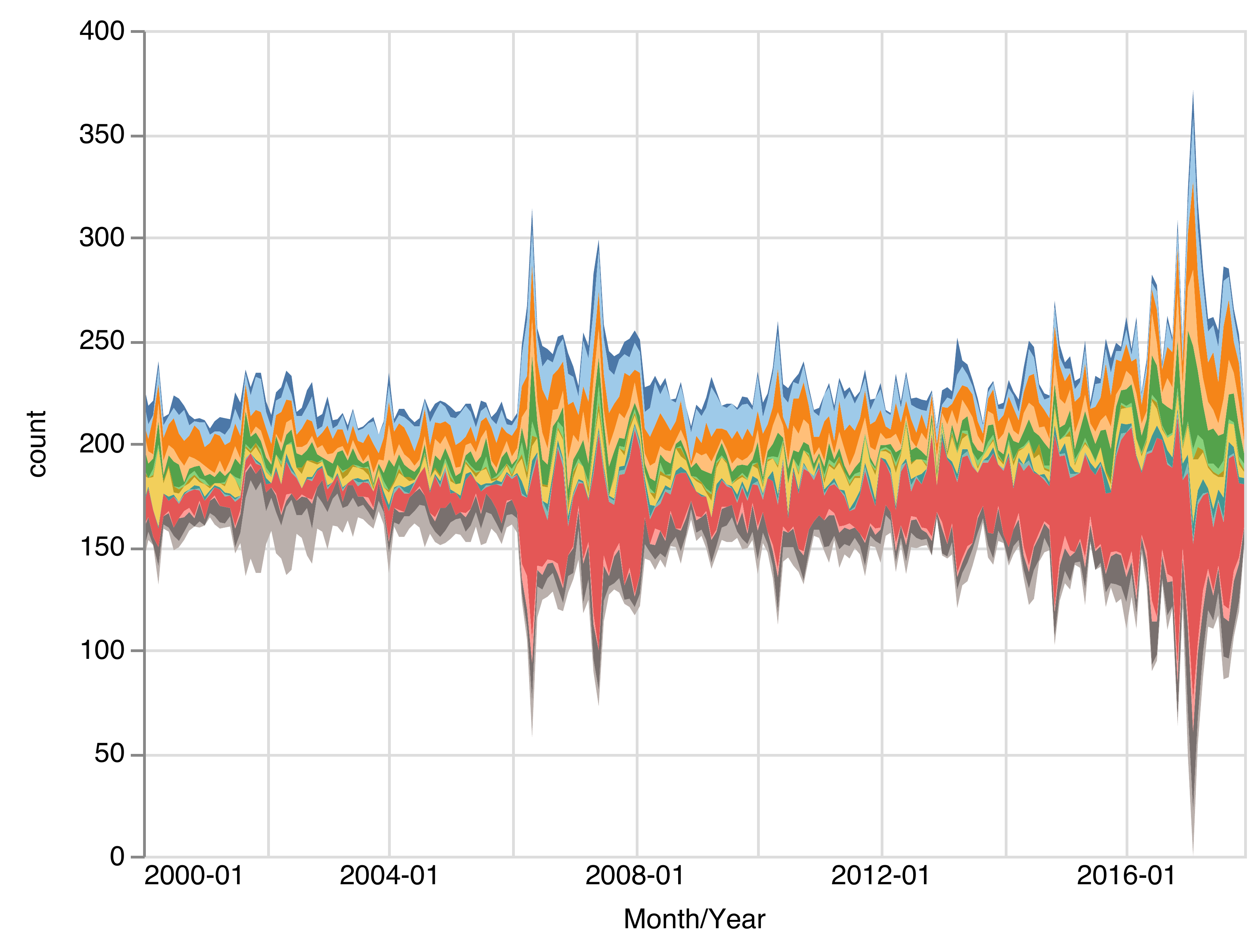}\label{fig:immigration}}
\subfigure[Football]{\includegraphics[width=0.23\textwidth]{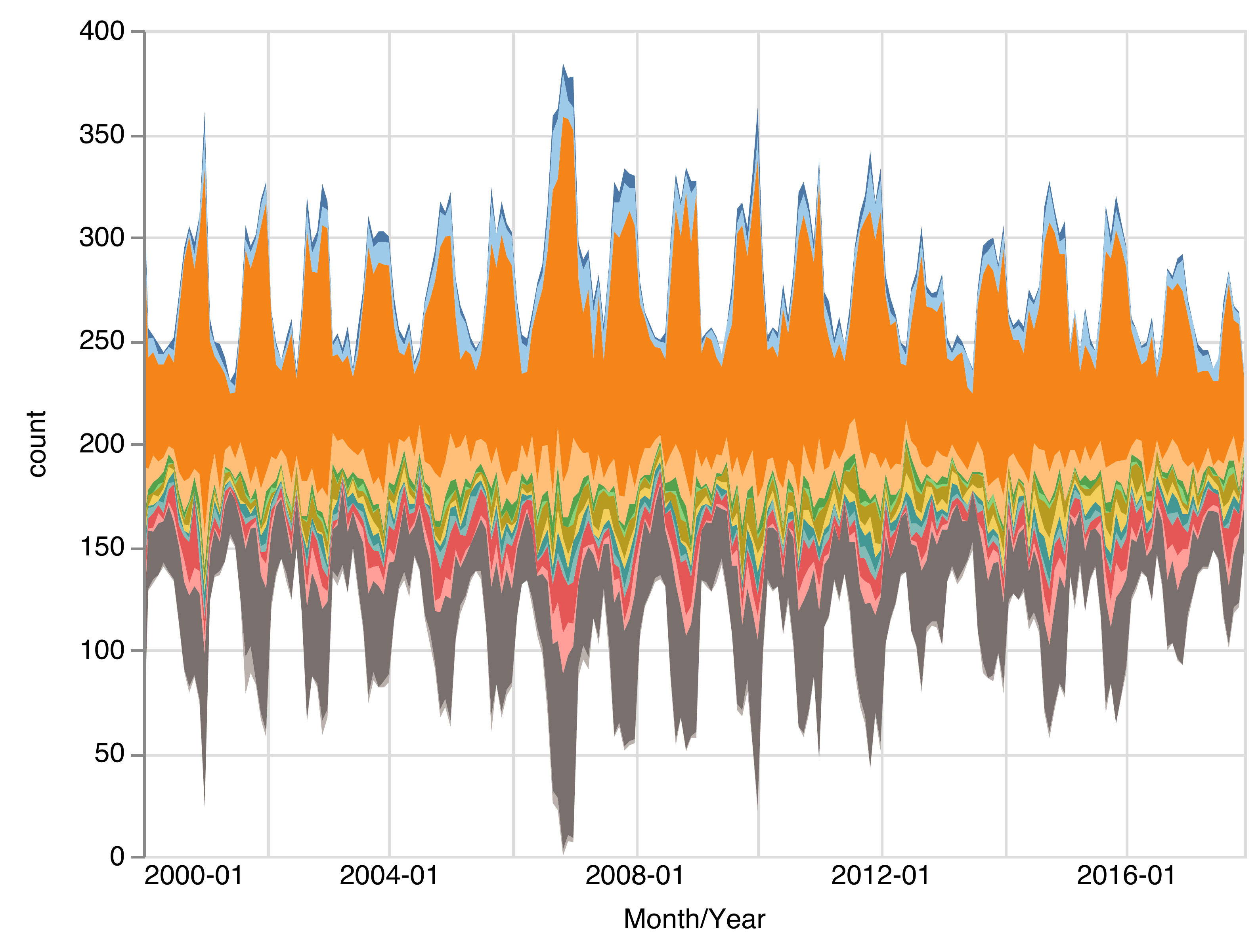}\label{fig:football}}
\subfigure[Hurricane]{\includegraphics[width=0.23\textwidth]{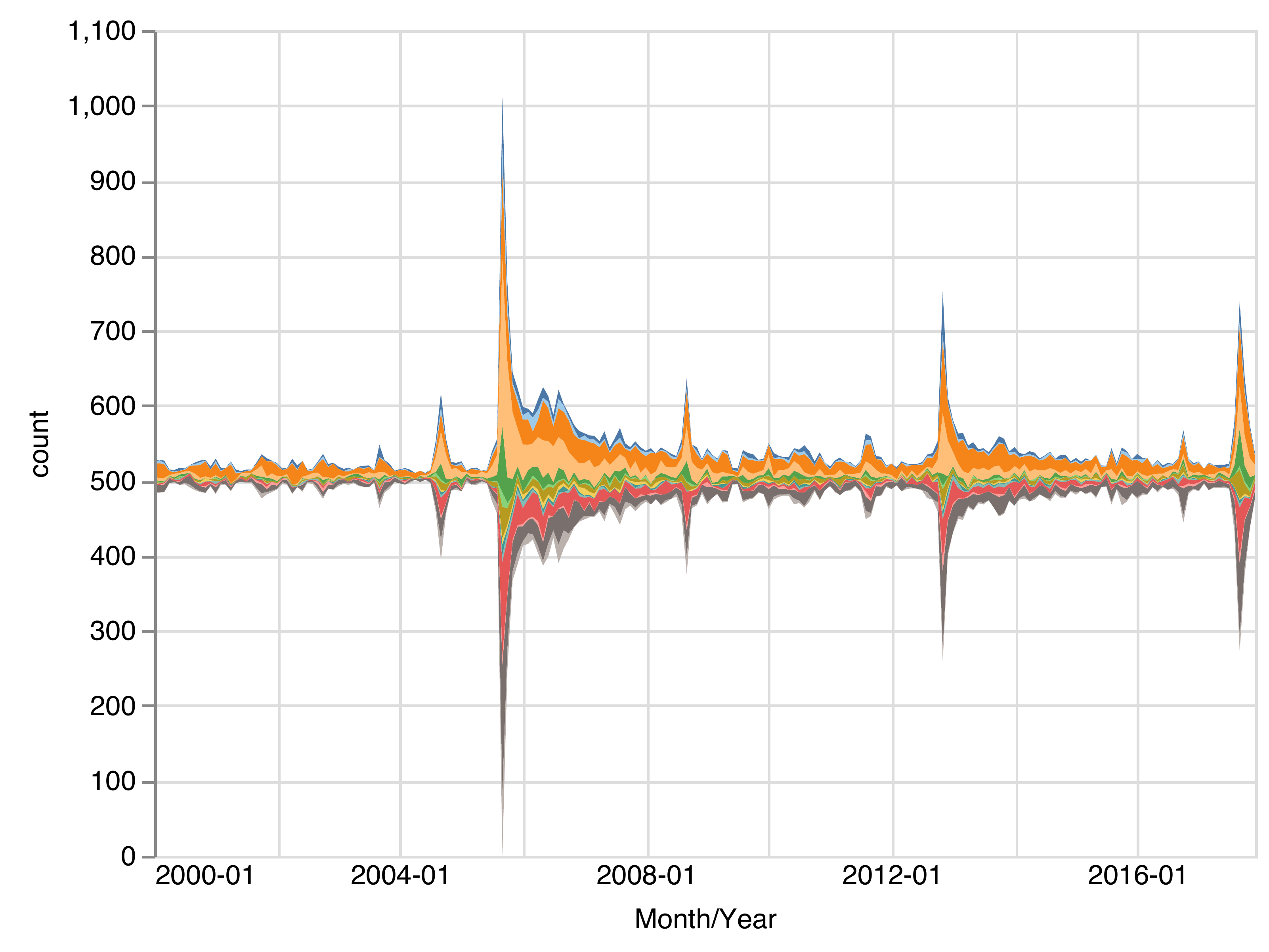}\label{fig:hurricane}}
\subfigure[Shooting]{\includegraphics[width=0.23\textwidth]{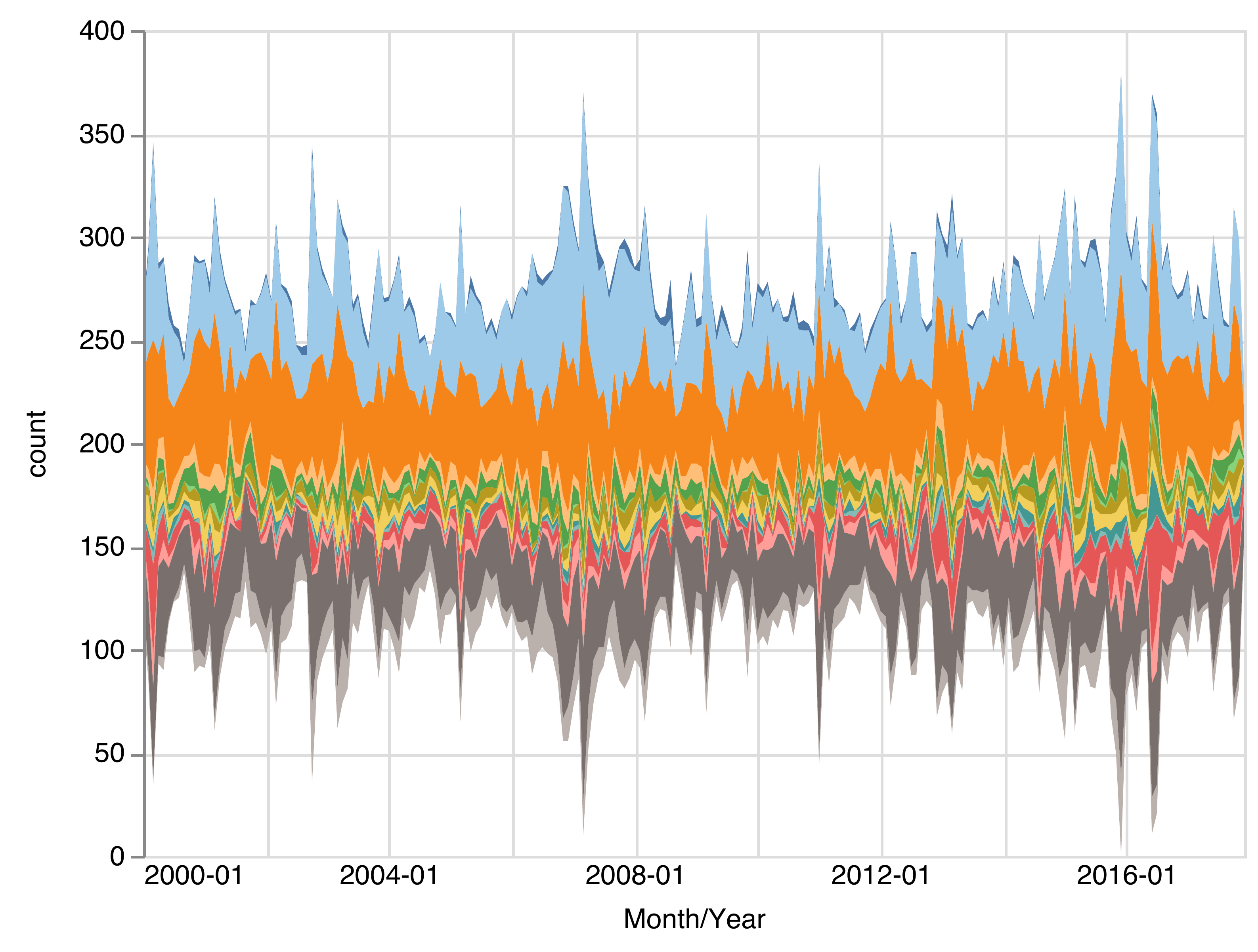}\label{fig:shooting}}

\caption{Frames by different issues}
\label{fig:frames_by_issues}
\end{figure*}

\section{Issue-level Framing}

We now move on to our second research question: \textit{RQ2: What are the framing patterns at the issue level?} 
While the New York Times shows a fairly stable temporal pattern of frame abundance with some exceptions, specific issues may not be stable at all.  Our question, for instance, in relation to the topic of abortion, whether there has been a consistent framing.

To quantitatively estimate the temporal change in issue-level framing, we use keyword matching to collect news articles around a particular issue and examine the temporal evolution of prominent frames. 
Two different types of issues are considered: 1) policies (e.g., `Abortion', `Smoking', and `Immigration') and 2) events and crises (e.g., `Football', `Hurricane', and `Shooting').  
These issues are chosen as illustrative examples to show different patterns of frame dynamics; similar dynamics are shared across other issues.

Figure~\ref{fig:frames_by_issues} shows a stream graph of the total counts of news articles around various issues over time (at the month level), broken down into the different media frames. 
For `Abortion' (Fig.\ref{fig:abortion}), we find that the Political and Legality frames dominate, while the presence of Morality is more  noticeable compared with other issues. 
For `Smoking' (Fig.\ref{fig:smoking}), we see that the Cultural, Quality of life, and Health frames  are consistently more prevalent than other frames. We also see that the Policy and External regulation  frames show a sudden increase in 2003, which may be due to New York City's smoke-free law that banned smoking in almost all  restaurants and bars in 2003.
For `Immigration' (Fig.\ref{fig:immigration}), we see the Security frame increasing in prevalence in 2001 after the September 11 attacks, and the prevalence of the Political frame increasing around the 2006 midterm elections. This suggests that the discussion on immigration tended to be increasingly associated with security after  September 11, and became one of the key issues in  subsequent elections. 

For `Football' (Fig.\ref{fig:football}), we see a cyclic pattern with a recent decreasing trend that synchronizes with the playing  seasons. The two frames relating to entertainment, the Cultural and Quality of life frames, have the strongest prevalence, as expected.
For `Hurricane' (Fig.\ref{fig:hurricane}), we see sporadic surges in the volume of news articles as hurricanes hit the U.S. 
Generally, the Economic frame seems to be the most prominent in the early stage of hurricanes, prevalence gradually decreasing, while the Cultural frame appears consistent over time. The importance shown by the Economic frame corroborates the results of previous studies~\cite{houston2012disaster}.

Lastly, `Shooting' has consistently received great attention with a few surges around mass shooting incidents (Fig.\ref{fig:shooting}). The most prominent frames are the Crime, Cultural, and Quality of life frames, but in recent years, the Political and Legality frames increased in prevalence, probably due to the more active public discussion of gun control laws. We investigate this phenomenon more closely in Section 7.

The results indicate that the eminence profile of general frames can be used as a signature for characterizing events or issues. 
Moreover, tracking the most prominent frames on a particular issue can reveal how the focus of the media changes over time.

\subsection{Framing in Issue Development}
\label{subsec:issue_development}

We now focus on the shorter time span and track how issues develop  from the perspective of media frames. 
Are there any framing patterns moving between the early, mid, and late stages of issue development? 
To answer this question, we focus on news articles about two mass shootings (in Orlando in 2016 and Las Vegas in 2017) and two hurricanes (Katrina and Sandy), which received much attention from news media.
We extract the news articles by keyword matching and limit results to those published within four weeks from the event (for the  hurricane event dates, we used 2005/08/23 for Katrina and 2012/10/22 for Sandy, i.e., the dates the hurricanes were formed). 
The four-week window was chosen based on previous studies of media coverage of mass shootings~\cite{muschert2006media,chyi2004media,muschert2006media}. 
Compared to mass shootings, natural disasters produce a much more diverse news lifespan, even lasting multiple years~\cite{houston2012disaster}. 
We consider a month as a long-enough time window to capture the most fundamental news cycle and levels of public attention. 
We then divide the news articles into three groups, which relate to  the early, mid, and late stages of issue development, respectively, based on publication dates.

We find diverse framing patterns along with the issue development. 
Hurricane Katrina, the costliest hurricane in U.S.  history, was described most commonly with the Economic frame, while Hurricane Sandy was reported most commonly with the Cultural identity frame in the early and late stages, with the Quality of life frame prevailing in the mid stage. 
The Orlando nightclub shooting, the deadliest incident of violence against LGBT people in U.S. history, was reported most commonly with the Morality frame in the early stage and with the Political frame in the late stage.  The Las Vegas shooting was reported with the Crime and punishment frame in the early stage and the Cultural identity and Quality of life frames in the later stages. 

We, however, consistently find \emph{framing convergence} across the different issues, meaning that the dominance of the most dominant frame at the late stage is stronger than that of its  counterpart at the early stage. 
For example, while the prevalence of the Morality frame in the early stage of the Orlando shooting coverage was 18.6\%, that of the Political frame in the late stage reached 40.7\%.  Similarly, the prevalence of the Economic frame increased from 21.4\% at the early stage to 27.5\% at the late stage in the coverage of Hurricane Katrina. 
This tendency is related to  frame-changing~\cite{muschert2009frame,chyi2004media}. To keep an  issue fresh to readers, journalists shift frames over time within an issue-attention cycle. For example, the emphasis of the news coverage of the Columbine school shooting moved from a focus on personal details to one on societal problems over the course of a  month~\cite{chyi2004media}, which is somewhat aligned with the framing convergence from the Mortality to the Political frame in relation to the Orlando shooting.

The framing convergence can be driven by a combination of multiple potential mechanisms.
At the early stage, there are lots of unknowns. News media need to report what is happening at the present moment from multiple perspectives. Moreover, the number of articles is high. 
Media are thus simply obligated to use diverse frames to report an issue at the early stage. 
Actual convergence on the importance of certain aspects may then take place in social discourses. 
While various perspectives are reported at the early stage, as time goes by, the public and the news media may form stronger consensus on the most important perspectives~\cite{zhou2007parsing}. 

\section{Frames and Delivery}

Here, we investigate our next research question: \textit{RQ3: How is each frame delivered? Is there a specific linguistic style (e.g., sentiment)?} 
Intuitively, some media frames might be likely to seek to evoke  a certain sentiment. For example, news articles with the Crime and punishment frame may be typically written with a negative sentiment.  
As the sentiment of the news articles can influence user perception, news reading behavior, and propagation dynamics~\cite{reis2015breaking}, understanding the association between media frames and their sentiment helps to assess the impact of these frames.
We measure sentiment of each news article by using the widely adopted VADER Sentiment Analyzer~\cite{bird2004nltk}. 
While the sentiment of a word can be changed by the context~\cite{an2018semaxis}, a lexicon-based method has been validated with a variety of massive text data, such as books~\cite{reagan2016emotional}, chats~\cite{kang2017would}, song lyrics~\cite{dodds2010measuring}, news articles~\cite{reis2015breaking}, and tweets~\cite{frank2013happiness}.
It should be noted that the sentiment of a headline can be taken as a proxy for the key sentiment of the corresponding article, because headlines are expected to express a concise summary of the article~\cite{bell1991language}.

\begin{figure}[ht]
\centering
\includegraphics[width=\columnwidth]{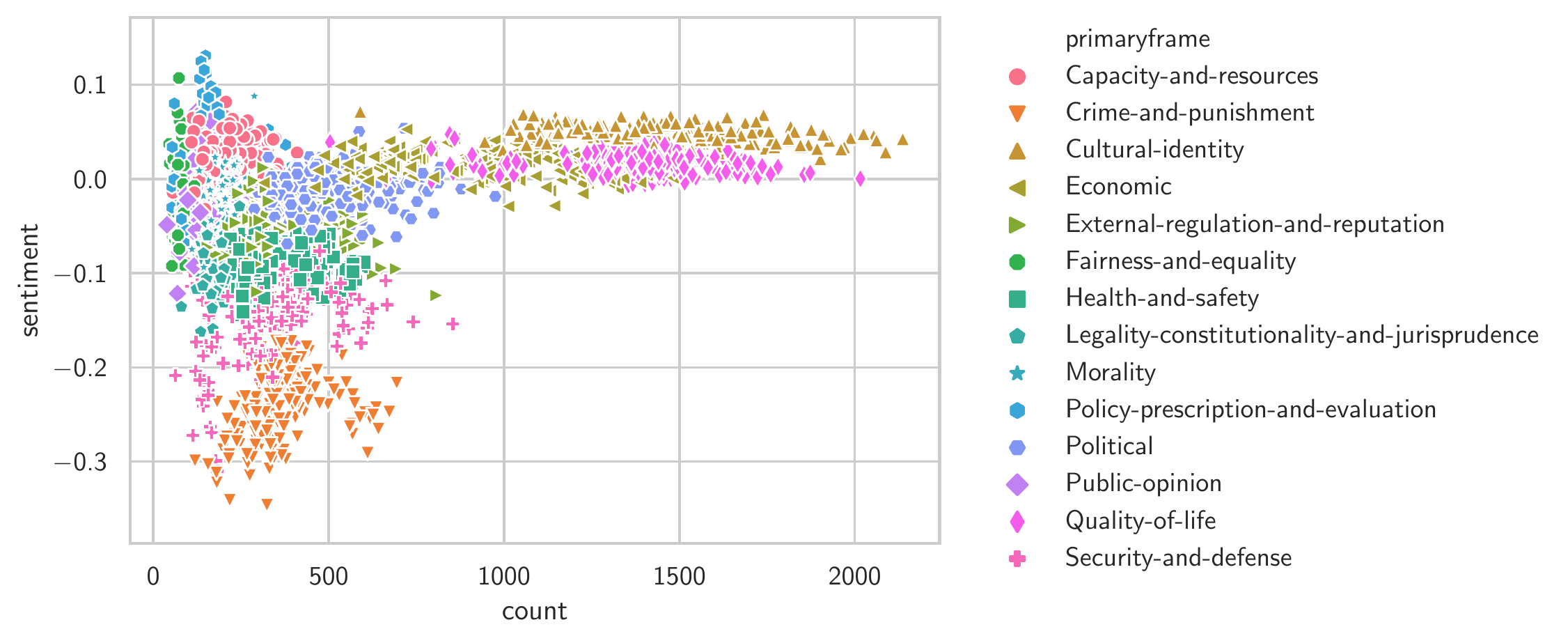}
\caption{Sentiments and frame (each dot represents a monthly average sentiment)}
\label{fig:tones_per_frame}
\end{figure}

Figure~\ref{fig:tones_per_frame} shows the number of articles with each frame (x-axis) and their average sentiments (y-axis), which are computed monthly.
The clustering of the dots from the same frame is visible. 
In other words, each media frame is delivered with a particular sentiment, and such associations are quite stable across different events occurring in the last 18 years.

\begin{figure}[ht!]
\centering
\includegraphics[width=\columnwidth]{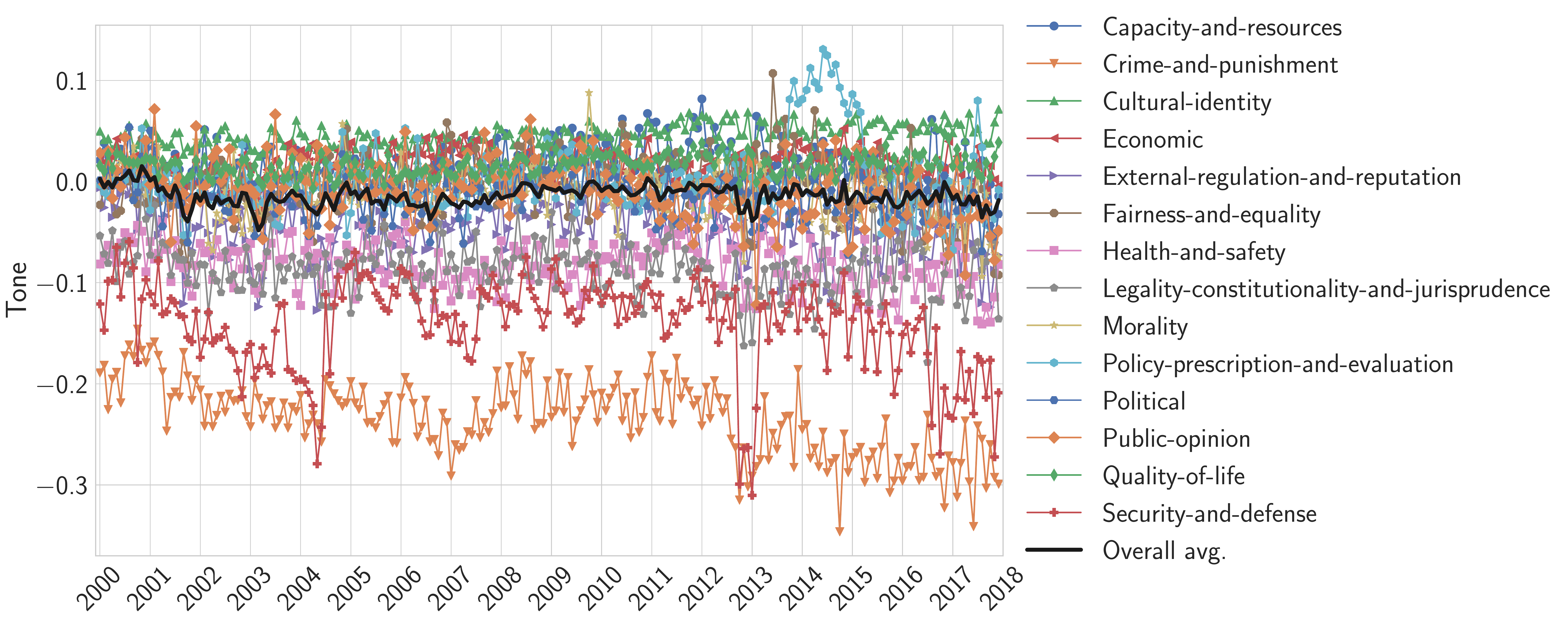}
\caption{Sentiments over time}
\label{fig:tones_over_time}
\end{figure}

Figure~\ref{fig:tones_over_time} shows the monthly average sentiments of the news articles for each frame over time. 
As the closely clustered dots for each frame in  Figure~\ref{fig:tones_per_frame} imply, in general, the sentiments associated with each frame are stable (average variation$<$0.1) over time.  
The solid black line represents the monthly average sentiment. Interestingly, the line is close to zero with a slightly negative bias, meaning that the overall sentiment of the news articles published monthly is almost neutral but slightly negative. 

\begin{figure}[ht]
\centering
\includegraphics[width=\columnwidth]{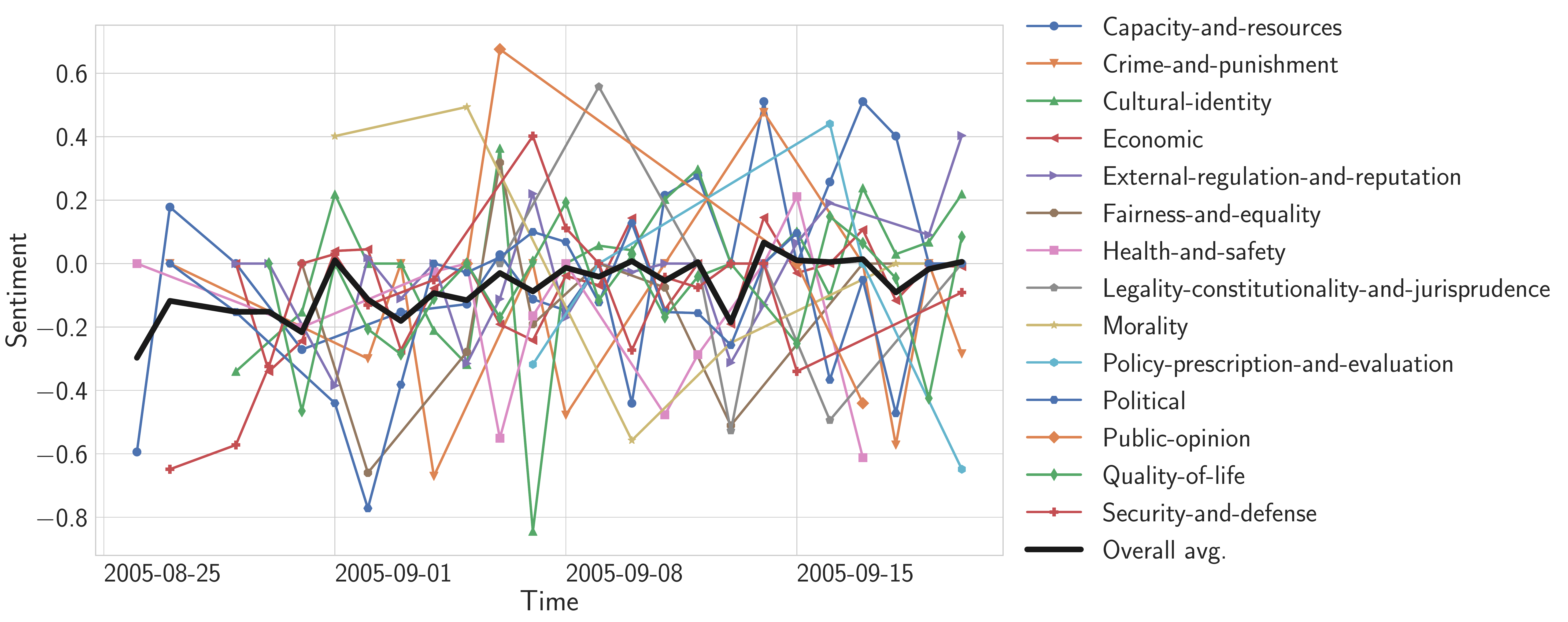}
\caption{Sentiments of news on Hurricane Katrina (2005)}
\label{fig:katrina}
\end{figure}

\begin{figure}[ht]
\centering
\includegraphics[width=\columnwidth]{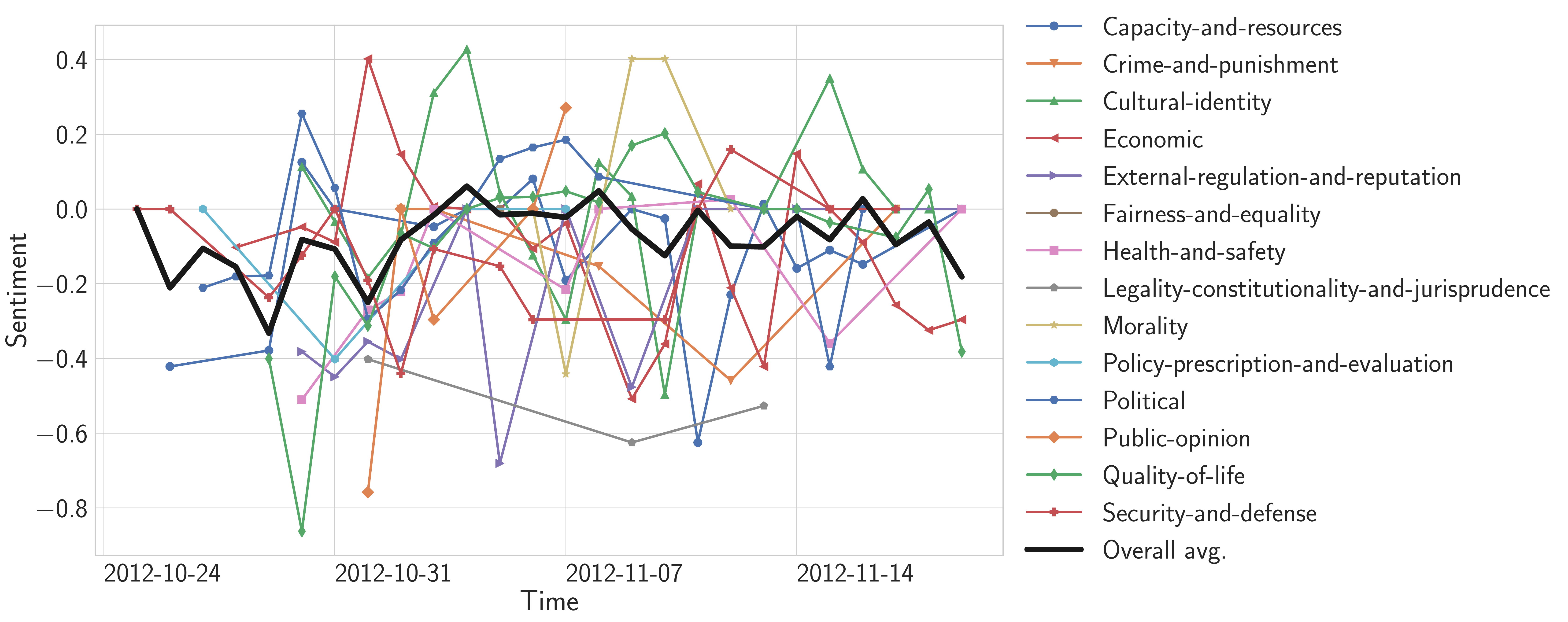}
\caption{Hurricane Sandy (2012)}
\label{fig:sandy}
\end{figure}

More interestingly, the aggregated neutral sentiment is also observed in a set of articles on specific issues that could be expected to have negative sentiment. Figures~\ref{fig:katrina} and \ref{fig:sandy} show the sentiment of the frames, as calculated from news articles relating to Hurricanes Katrina and Sandy, respectively. 
We extract news articles by using the keywords `Hurricane Katrina' and `Hurricane Sandy', and limit the articles to those published in the month after the hurricanes formed. 

In contrast to our expectation that disaster-related news be  written with a negative sentiment, the aggregated sentiment is, to some extent, neutralized by positive news.
For example, news about aid or relief can be delivered through various frames, including the Economic or Capacity and resources frames. Or, dramatic and touching human-interest stories are delivered through the Cultural identity frame. 
Thus, even disaster-related news does not solely convey a negative mood but tends to be balanced.
While it is debatable whether this `balanced' sentiment is an intentional consequence of editorial decisions~\cite{galician1987balancing}, we would note that, in spite of some media frames that are typically written in a negative tone, the aggregated sentiment of temporally or issue-specific news articles can be neutral.

\section{Cast Study: Framing of Mass Shootings}

In this section, we focus on mass shootings in the United States as a case study to demonstrate how issue-agnostic frames can help us understand the differences in framing among similar events. 
In contrast to the previous media studies on disasters~\cite{houston2012disaster} or shootings~\cite{muschert2006media}, which relied on manual coding, we demonstrate the potential of large-scale media frame research.
In Section~\ref{subsec:issue_development}, we have shown that the media coverage of the two mass shootings went through different development patterns in terms of the issue framing.  We now extend this analysis on a more systematic footing.

\begin{table}[ht]
\footnotesize
\begin{center}
\begin{tabular}{c|c|c}
\toprule
Incident & Date & Deaths  \\
\midrule
Virginia Tech shooting & 2007/4/16 & 32 \\
Geneva County massacre & 2009/3/10 & 10  \\
Binghamton shootings & 2009/4/3 &	13  \\
Fort Hood shooting & 2009/11/5 & 14  \\
Aurora shooting	& 2012/7/20 &	12	 \\
Sandy Hook School shooting &	2012/12/14 &	27 \\
Washington Navy Yard shooting &	2013/9/16 &	12   \\
San Bernardino attack &	2015/12/2 &	14  \\
Orlando nightclub shooting &	2016/6/12 &	49   \\
Las Vegas shooting &	2017/10/1 &	58  \\
Sutherland Springs church shooting & 2017/11/5 &	26  \\
\bottomrule
\end{tabular}
\end{center}
\caption{Mass shootings occurred during 2000 to 2017~\cite{mass_shooting}}
\label{tbl:mass_shootings}
\end{table}

We first compile a list of 11 mass shootings that occurred in the U.S. from 2000 to 2017~\cite{mass_shooting}, shown in Table~\ref{tbl:mass_shootings}. We then extract articles appearing during the month after the incident using `shooting' and its location as keywords. One month is the most commonly used time window for studying the media coverage of shootings~\cite{muschert2006media,chyi2004media,muschert2006media}.  
For each of the 11 shootings, we compute the proportion of each media frame and construct a 14-dimensional vector whose dimension maps onto the corresponding frame.

\begin{figure}[ht]
\centering
\includegraphics[width=\columnwidth]{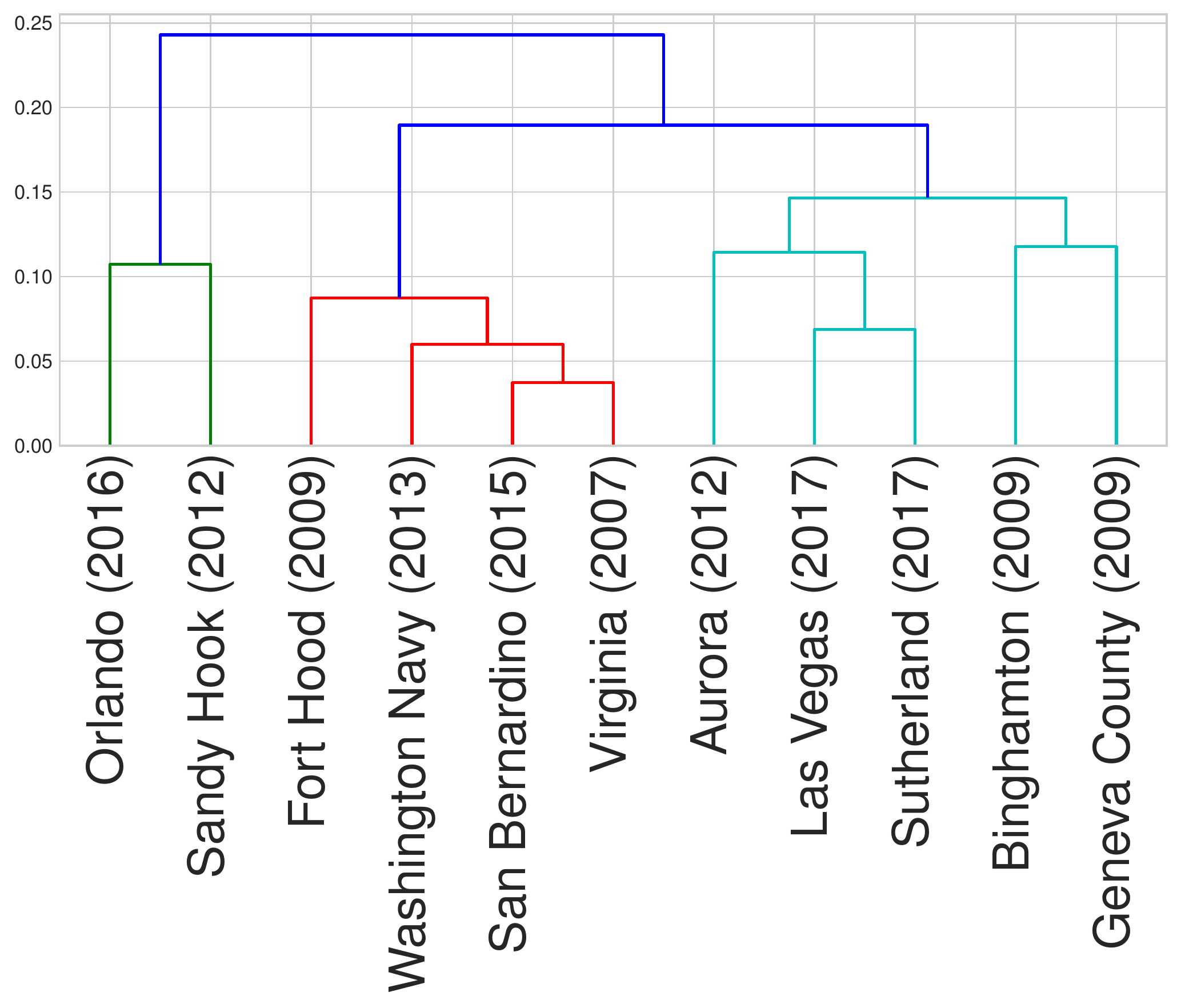}
\caption{Clusters of 11 mass shootings}
\label{fig:shooting_cluster_dendrogram}
\end{figure}

Figure~\ref{fig:shooting_cluster_dendrogram} shows the clusters of mass shootings by hierarchical clustering of mass shooting vectors. 
We use Euclidean distance as the distance metric and the Ward variance minimization algorithm as the linkage method. 
We refer to the green cluster (which covers the Orlando nightclub shooting and the Sandy Hook Elementary School shooting) as  $C_g$, the red cluster (which covers the Fort Hood shooting, the Washington Navy Yard shooting, the San Bernardino attack, and the Virginia Tech shooting) as $C_r$, and the sky-blue cluster (which covers the rest of the shootings) as $C_s$.
The average prevalences of each media frame for $C_g$, $C_r$, and $C_s$ are presented in Table~\ref{tbl:shooting_cluster}. 
We also mark the discriminating frames if the average proportion of a particular frame in one cluster is one standard deviation away (either above (up arrow $\uparrow$) or below (down arrow $\downarrow$)) from that in the other two clusters.

\begin{table}[ht]
\footnotesize
\begin{center}
\begin{tabular}{c|c|c|c}
\toprule
Frame & $C_{g}$ & $C_{r}$ & $C_{s}$ \\
\midrule
Capacity and resources & 0.011 & \textbf{0.014}$\uparrow$ & 0.005 \\
Crime and punishment & \textbf{0.166}$\downarrow$ & \textbf{0.241}$\uparrow$ & 0.204 \\
Cultural identity & \textbf{0.18}$\downarrow$ & 0.202 & \textbf{0.24}$\uparrow$ \\
Economic & 0.034 & 0.029 & \textbf{0.045}$\uparrow$ \\
External regulation & 0.035 & 0.031 & \textbf{0.043}$\uparrow$ \\
Fairness and equality & 0.007 & 0.008 & 0.009 \\
Health and safety & \textbf{0.059}$\uparrow$ & 0.046 & \textbf{0.038}$\downarrow$ \\
Legality constitutionality & 0.024 & 0.023 & 0.021 \\
Morality & \textbf{0.051}$\uparrow$ & 0.017 & 0.025 \\
Policy prescription & 0.022 & 0.012 & 0.01 \\
Political & \textbf{0.163}$\uparrow$ & 0.058 & 0.06 \\
Public opinion & \textbf{0.015}$\downarrow$ & 0.028 & \textbf{0.034}$\uparrow$ \\
Quality of life & 0.189 & 0.191 & \textbf{0.225}$\uparrow$ \\
Security and defense & 0.045 & \textbf{0.1}$\uparrow$ & \textbf{0.041}$\downarrow$ \\
\bottomrule
\end{tabular}
\end{center}
\caption{Average prevalences of each media frame for mass shooting clusters.}
\label{tbl:shooting_cluster}
\end{table}

The discriminating frames show which perspectives were more emphasized in the reporting of each cluster of mass shootings. 
We find that the two shootings ($C_g$), whose victims were minority  groups, tend to produce more articles on public safety, morality, and political impact. 
Considering that the Sandy Hook Elementary School shooting provoked a national debate on gun control and the Orlando gay nightclub shooting occurred during the final stages of the 2016 United States presidential election campaign, it is understandable that the Political frame is particularly prominent.
The four shootings ($C_r$), respectively at the military base (Fort Hood), the Naval Sea Systems Command (Washington Navy), the government-funded public benefit corporation (San Bernardino), and the public university (Virginia), are reported with a stress on crime and threats to welfare.  
The remaining five mass shootings in $C_s$ show a higher prevalence  of the Cultural identity and Quality of life frames than the other two clusters. The Cultural identity frame is used when explaining the motives of suspects, and the Quality of life frame is used for a human-interest story.  

The discriminating frames with a down arrow in Table~\ref{tbl:shooting_cluster} can be understood to represent the news stories that were not published. Since the number of articles published per day is still limited, even in the online era, newsrooms may first develop more critical stories and delay others  (and sometimes cancel them if the story loses news value due to delays).

\section{Discussion and Future Work}

This work has built a media frame classifier and used it to conduct a systematic media frame analysis of 1.5 million news articles from the New York Times published from 2000 to 2017. 
We fine-tuned the pre-trained BERT-base model using the Media Frames Corpus and achieved a higher performance than the best-existing method, which uses discourse structure and a recursive neural network.
We addressed  research questions about the long-term trends, issue-level framing dynamics, and delivery of frames in news reporting. We also showed that issue-agnostic media framing can be a helpful tool to differentiate articles about similar events (e.g., mass shootings).

We found that, although the most prominent media frames are fairly stable, there exist long-term trends of highly predictive fluctuation in relation to major social events. 
The gradual shift from economic- to culture- and quality of life-related frames could be explained by the societal trend toward more sophisticated cultural needs~\cite{grossberg2006mediamaking} and the influences of market forces~\cite{hamilton2004all} over time.
We also found that the eminence profile of topic-agnostic frames can be used as a signature for characterizing events or issues. By  digging into the framing patterns of each issue, we discovered frame convergence, whereby the dominance of the most dominant frame at the late stage is stronger than that of its counterpart at the early stage. This might reflect the many unknowns in the early stage and the consensus formed in the late stage.
Moreover, we studied the association between media frames and  average sentiment. We showed that even though some frames (e.g., Crime) are strongly connected to a specific one (e.g., negative), the aggregate tone across the time and topics is almost neutral. 
By delving into the news coverage of mass shootings in the United States from 2000 to 2017, we revealed three clusters of framing patterns organized primarily: around who the victims were, where the shooting occurred, and the motives of the suspect(s). 

For future work, 
we consider several research directions.
First, we aim to leverage the multilingual word embeddings to port our classifier to non-English news articles. In these contexts, media frame classification is often conducted by extracting representative English words from the MFC, translating the word into the target language, and extending the lexicons~\cite{field2018framing}. We will explore the potential of the multilingual word embeddings to simplify the process of  detecting media frames in non-English news articles. Second, we will extend our analysis to the comparison of multiple news media outlets.
Although the New York Times is an authoritative source for agenda-setting and plays a critical role in shaping public opinion, comparing various news media outlets from the perspective of their prominent frames on different issues will allow a more comprehensive picture. 
For example, it would be interesting to examine the political leanings of the news media in terms of the dominant frames on certain issues. 
Last but not least, with the advancement of neural language models, framing detection will become more accurate and be applied to more and more diverse text data. As the general media frames are universal across different topics, they may serve as a useful means of categorizing social media posts (e.g., tweets) across issues, providing a new tool to study the interaction between media and the public.

\begin{acks}
We would like to thank the anonymous reviewers who helped to improve the manuscript. Y.Y.A. is supported by the Defense Advanced Research Projects Agency (DARPA), contract W911NF-17-C-0094. The U.S. Government is authorized to reproduce and distribute reprints for Governmental purposes notwithstanding any copyright annotation thereon. The views and conclusions contained herein are those of the authors and should not be interpreted as necessarily representing the official policies or endorsements, either expressed or implied, of DARPA or the U.S. Government.
\end{acks}

\balance
\bibliographystyle{ACM-Reference-Format}
\bibliography{kwak122_arxiv}




\end{document}